\documentclass[10pt]{article}

\usepackage{graphicx}
\usepackage{xcolor}
\usepackage{amsmath}
\usepackage{amssymb}
\usepackage[noabbrev]{cleveref}
\usepackage{cite}
\usepackage{lineno}

\usepackage[left=2.5cm,top=3cm,right=2.5cm,bottom=3cm,bindingoffset=0.5cm]{geometry}

\newcommand{\p}{\partial}
\newcommand{\f}{\frac}

\begin{document}
{\Large \bf \flushleft Effect of radial pressure corrugations and profile shearing on \\ turbulence in Fusion plasmas}\\[3ex]
{Ajay C. J$^1$, M. J. Pueschel$^{1,2}$, Justin Ball$^3$, David Hatch$^4$, Tobias G{\"o}rler$^5$ and Stephan Brunner$^3$}\\[3ex]
$^1$Dutch Institute for Fundamental Energy Research, 5612 AJ Eindhoven, The Netherlands\\
$^2$Eindhoven University of Technology, 5600 MB Eindhoven, The Netherlands \\
$^3$Ecole Polytechnique Fédérale de Lausanne (EPFL), Swiss Plasma Center (SPC), CH-1015 Lausanne, Switzerland\\
$^4$Institute for Fusion Studies, University of Texas at Austin, Austin, TX 78712, United States of America\\
$^5$Max Planck Institute for Plasma Physics, Boltzmannstr. 2, 85748 Garching, Germany
\\[3ex]
trax.42@hotmail.com
\\[2ex]
July 2025\\[4ex]

\begin{abstract}
Microturbulence can produce stationary fine-scale radial corrugations on the plasma density and temperature gradients in magnetic confinement fusion devices. 
We show that these structures play a significant role in regulating turbulent transport.
We focus on the pedestal, studying electron-temperature-gradient (ETG) mode destabilisation and saturation in the presence of radial corrugations on the electron temperature gradient that could result from microtearing turbulence.
A linear dispersion relation is derived for a shearless slab case, which indicates that in the presence of a sinusoidal background corrugation, each ETG mode splits into three distinct eigenvalues, with one being the original, one being more unstable and one being less unstable. However, despite the presence of more unstable linear modes, nonlinear gyrokinetic simulations of ETG with corrugated background electron temperature show a reduction of fluxes. Our investigation reveals a radial variation of the phase velocity of the modes that is proportional to the diamagnetic drift velocity and the local pressure gradient. The associated  profile shearing
breaks the turbulent eddies apart, reducing the transport level. 
This profile shearing resulting from fine-scale pressure corrugations could be a ubiquitous turbulence saturation mechanism not just in Fusion plasmas, but in Astrophysics and other areas.
\end{abstract}

\clearpage

\vspace{8ex}
\section{Introduction}\hfill

Loss of heat and particles due to microturbulence is one of the main challenges in maintaining reactor relevant conditions in magnetic confinement fusion devices such as tokamaks and stellarators. Plasma microinstabilities such as Ion temperature Gradient (ITG), Electron Temperature Gradient (ETG), Trapped Electron Mode, microtearing mode etc., all driven unstable by the radial gradient in background pressure profile, are some of the common instabilities contributing to turbulent transport~\cite{Hazeltine1975,Horton1999}.

Typically in studying such instabilities, both theoretically and using local numerical simulations, a constant gradient in pressure profile is assumed across the radius. However, studies~\cite{Waltz2006,Dominski2015,AjayCJ2020,Hatch2021,AjayCJ2023} have revealed ion Larmor radius scale corrugations on the pressure profile caused by these microinstabilities themselves. {\color{black}Investigating the impact of such fine-scale corrugations on the tertiary growth of instabilities and their saturation therefore becomes relevant. While there are related studies, they are few. 
The excitation of subcritical KBMs~\cite{Waltz2010,Pueschel2013_3}, the effect of second derivative of flow shear on L-H transition~\cite{Terry1994}, Dimits shift explanation based on second derivative of zonal velocity~\cite{Zhu2020}, effect of non-uniform radial variation of safety factor~\cite{Justin2022}, applying global profile curvature to local gyrokinetics~\cite{Candy2025} etc. are a few examples.

In this paper, we investigate the effect of background pressure corrugations on ETG (and ITG) modes. Microtearing~\cite{Pueschel2020,HamedPhD,Hatch2021} and ETG modes~\cite{Nevins2006,Parisi2020,Guttenfelder2021,Chapman2022} contribute towards turbulent transport in the pedestal~\cite{Kotschenreuther2019}. Previous studies~\cite{Pueschel2020} have explored ETG-microtearing multiscale effects. Here, we focus on a specific problem that has become relevant in the light of recent findings~\cite{Hatch2021,AjayCJ2023}, where microtearing modes are shown to produce corrugations whose gradients are comparable in amplitude to the background gradient, such that the electron temperature is almost fully flattened over a radial width of $1-10$ ion Larmor radii near the rational surfaces. We investigate ETG mode evolution and saturation in the presence of fixed externally imposed corrugations on the background electron temperature, similar to those caused by microtearing modes. 

A local dispersion relation is derived for a shearless slab system to include a sinusoidal variation in the background temperature gradient. The result shows that, in the presence of the corrugation, each eigenmode splits into three, with one being more unstable and one being less unstable than the original. Linear gyrokinetic simulations confirm that the linear growth rate increases with increasing amplitude of corrugations.

However, despite the presence of more unstable modes, nonlinear gyrokinetic simulations predict less turbulent flux in the presence of corrugations. Our analysis reveals that the phase velocity of the modes along the toroidal/poloidal direction varies with the radius, following the corrugation in the background pressure gradient and the associated diamagnetic drift velocity. The resulting `profile shearing'~\cite{Waltz2002,Ben2010} breaks apart the turbulent eddies and therefore lowering transport. In normal gyrokinetic simulations where such Larmor radius scale corrugations in the background pressure profiles are usually not included, transport may hence be over-estimated.

This paper is organised as follows. In section~\ref{SecDR}, the derivation of the slab dispersion relation and its results are presented. In section~\ref{SecGK}, the gyrokinetic simulation results are shown. First, a slab linear ETG case is considered in subsection~\ref{SecGENESlab} to verify the slab dispersion relation results against gyrokinetic simulations, followed by toroidal linear ETG results in subsection~\ref{SecLin}. The nonlinear simulation results are presented in subsection~\ref{SecNonlin}, where the profile shearing effect is discussed. Finally, in section~\ref{SecConc}, the conclusions are drawn.

\section{Dispersion relation for a shearless slab with background corrugations}\label{SecDR}\hfill

Dispersion relations for plasmas can be obtained by solving the Vlasov-Maxwell system of equations under various limits. Usually, they are derived under the assumption of constant backgrounds and gradients. Here, a local dispersion relation for a shearless slab system with radial corrugations on the background pressure is derived. In addition to being more analytically tractable than complex geometries, the choice of a slab system is motivated by the presence of slab-like ETGs that have been reported in the pedestal~\cite{Parisi2020,Hassan2021,Chapman2022}. {\color{black}Note that, in a slab system, the resonance of the wave with the parallel streaming of particles is relevant for instability, while in a toroidal system, the resonance with the curvature and $\nabla B$ drifts tends to be more relevant.}

The derivation of the dispersion relation is presented in subsection~\ref{SecDRDer}, followed by the results in subsection~\ref{SecDRDerResult}.

\subsection{Dispersion relation derivation}\label{SecDRDer}\hfill

The derivation in section 8.2 of reference~\cite{Ichimaru1973} and section 1.2 of reference~\cite{StephanMicroinstabilities}, which considers a constant background pressure gradient, is extended to include a radial cosine variation to the pressure gradient. {\color{black} First, the equilibrium set-up is described in subsection~\ref{SecDREq}. In subsection~\ref{SecDRdf}, the linearised Vlasov equation is solved for the general case of any periodic background pressure gradient profile and an expression for the perturbed distribution function is obtained in equation~(\ref{Eqdf}). In the next subsection~\ref{SecDRDief}, the dielectric function is obtained, whose zeroes give the modes of interest. Finally in subsection~\ref{SecDRCos}, the particular case of a cosine variation of the background pressure profile is considered, and the corresponding dielectric response is obtained in equation~(\ref{EqDRmat})}.

\subsubsection{Setting up the equilibrium.}\label{SecDREq}\hfill \vspace{1ex}

Cartesian coordinates $(x,y,z)$ are considered such that a uniform magnetic field $\vec{B}$ is aligned along $\hat{e}_z$, and variation in equilibrium pressure is along the $\hat{e}_x$ direction. {\color{black} No equilibrium electric field is considered.} A charged particle follows the magnetic field line along $\hat{e}_z$, while gyrating around it in the ($x,y$) plane with a gyrofrequency $\Omega$, while simultaneously seeing an equilibrium variation in $x$. $\Omega=qB/m$, where $q$ is the particle charge, $m$ is the mass and $B=|\vec{B}|$.

The equilibrium distribution function $f_0(X,\mathcal{E})$ of a given plasma species is a function of the constants of motion of the unperturbed system, and is assumed to be near-Maxwellian: 
\begin{align*}
f_0(X,\mathcal{E})=\frac{N(X)}{[2\pi T(X)/m]^{3/2}}e^{-\mathcal{E}/T(X)},
\end{align*}
where, the invariants $X=x+v_y/\Omega$ is the radial position of the gyrocenter of a particle, and $\mathcal{E}=mv^2/2$ is the kinetic energy. $N$ is the background density and $T$ is the background temperature. 

Assuming that the characteristic length $L$ of variations of the equilibrium profiles are much smaller than the Larmor radius $\rho$ such that $\epsilon=\rho/L \ll 1$, one can Taylor expand $f_0$ as:
\begin{align}
f_0(X,\mathcal{E}) 
&= f_0(x,\mathcal{E}) + \frac{\p f_0(x,\mathcal{E})}{\p x}\frac{v_y}{\Omega} + \mathcal{O}(\epsilon^2), {\rm where},
\label{eqf0}
\\
\frac{\p f_0}{\p x}
&=
\bigg( \frac{d {\rm ln} N}{dx} + \frac{dT}{dx}\frac{\p}{\p T}
\bigg) f_0. 
\label{eqdf0dx}
\end{align}

\subsubsection{Solving the linearised Vlasov equation.}\label{SecDRdf}\hfill \vspace{1ex} 

Let the perturbed components of the distribution function $\delta f$ ($ = f-f_0$) and the electrostatic potential $\delta \Phi$ be of the form: 
\begin{align}
\delta f &=\sum_{k_x}\hat{\delta f} e^{i(k_xx+k_yy+k_zz - \omega t)},
\label{EqfFourier}\\
\delta \Phi &=\sum_{k_x}\hat{\delta \Phi} e^{i(k_xx+k_yy+k_zz - \omega t)}.
\label{EqphiFourier}
\end{align}
Given that the unperturbed system is invariant along $y$ and $z$, and is stationary, $\delta f$ and $\delta \Phi$ have fixed wavenumbers $k_y$ and $k_z$, and fixed frequency $\omega \in \mathbb{C}$. The perturbation $\delta f$ is the solution of the linearised Vlasov equation:
\begin{align*}
\frac{D}{Dt}\bigg\vert_{u.t.}\delta f
= \bigg[ \frac{\p}{\p t}
+ \vec{v}\cdot \frac{\p}{\p \vec{r}}
+ \frac{q}{m}(\vec{v} \times \vec{B}) \cdot \frac{\p}{\p \vec{v}}
\bigg] \delta f
=\frac{q}{m} \frac{\p \delta \Phi}{\p\vec{r}} \cdot\frac{\p f_0}{\p \vec{v}
},
\end{align*}
where $D/Dt\vert_{u.t.}$ is the total time derivative along the unperturbed trajectories of the particles. The solution to $\delta f$ can be obtained by integrating along the unperturbed trajectories:
\begin{align}
\delta f(\vec{r},\vec{v},t)
= \frac{q}{m} \int_{-\infty}^{t} dt' \frac{\p \delta \Phi}{\p\vec{r}}\cdot\frac{\p f_0}{\p \vec{v}}\bigg\vert_{\vec{r}',\vec{v}',t'},
\label{Eqfint}
\end{align}
with the condition that $\delta f(t=-\infty)=0$. The unperturbed trajectory $[\vec{r}',\vec{v}']$ of a particle correspond to the gyromotion of the particle along the magnetic field $\vec{B}$:
\begin{align}
\vec{r}'(t')
=& \vec{r} + \frac{1}{\Omega}{\bf Q}(\tau)\vec{v},
\label{Eqr}\\
\vec{v}'(t')
=& {\bf R}(\tau)\vec{v},
\end{align}
where $\tau=t'-t$, and matrices ${\bf Q}$ and ${\bf R}$ are:
\begin{align*}
{\bf Q}(\tau) 
=
\left(
\begin{matrix}
{\rm sin}(\Omega\tau)  &  -[{\rm cos}(\Omega\tau) - 1]  &  0\\
{\rm cos}(\Omega\tau) - 1  &  {\rm sin}(\Omega\tau)  &  0\\
0  &  0  &  \Omega\tau
\end{matrix}
\right)
\\
{\bf R}(\tau) 
=
\frac{1}{\Omega}\frac{d}{d\tau} {\bf Q}
=
\left(
\begin{matrix}
{\rm cos}(\Omega\tau)  &  {\rm sin}(\Omega\tau)  &  0\\
-{\rm sin}(\Omega\tau)  &  {\rm cos}(\Omega\tau)  &  0\\
0  &  0  &  1
\end{matrix}
\right).
\end{align*}

Making use of \cref{eqf0,eqdf0dx}, and $\p f_0(x,H)/\p \vec{v}=-(\vec{v}/v_{th}^2)f_0(x,H)$, one obtains:
\begin{align}
\frac{\p f_0}{\p\vec{v}}
= 
\bigg[ \frac{\hat{e}_y}{\Omega}
\bigg( \frac{d {\rm ln} N}{dx}
+ \frac{d T}{dx}\frac{\p}{\p T}
\bigg) - \frac{\vec{v}}{v_{th}^2}
\bigg] f_0,
\label{Eqdfdv}
\end{align}
correct to first order in $\epsilon$.

{\color{black}The background profile gradients in the previous relation providing the drive of the instability can be expressed in terms of the diamagnetic drift frequency $\omega_d$ as:}
\begin{align}
\omega_d(x)
=\sum_{k_x}\hat{\omega}_{d,k_x} e^{ik_xx}
=\frac{Tk_y}{qB}
\left(\frac{d{\rm ln} N}{dx} + \frac{dT}{dx}\frac{\p}{\p T}
\right).
\label{Eqwd}
\end{align}
{\color{black}Unlike in the original derivation~\cite{Ichimaru1973} where $\omega_d$ is considered to be a constant, here $\omega_d(x)$ is assumed to be a continuous periodic function in $x$} in the long wavelength limit such that $k_x\rho\ll 1$. Note that, $\omega_d$ remains a differential operator with respect to $T$.

Using~\cref{EqfFourier,EqphiFourier,Eqfint,Eqdfdv,Eqwd}, one obtains:
\begin{align}
\hat{\delta f}_{k_x}
= \frac{q}{m}
\sum_{k'_x}\f{i}{v_{th}^2}
\left[
\hat{\omega}_{d,k_x-k'_x} - \vec{k}\cdot \vec{v}'\delta_{k_x,k'_x}
\right]
\hat{\delta \Phi}_{k'_x} \int_{-\infty}^{t} dt' e^{i(\vec{k}[\vec{r}'-\vec{r}] - \omega[t'-t])}.
\label{Eqf1}
\end{align}

To evaluate equation~(\ref{Eqf1}) and arrive at the final form in equation~(\ref{Eqdf}), the following three relations prove useful.
\begin{enumerate}
\item The Jacobi-Anger expansions:
\begin{align}
e^{iz\,{\rm sin}\theta}
&=\sum_{n=-\infty}^{\infty}J_n(z)e^{in\theta},
\label{Eqja1}\\
e^{iz\,{\rm cos}\theta}
&=\sum_{n=-\infty}^{\infty} i^n J_n(z)e^{in\theta}.
\label{Eqja2}
\end{align}

\item Making use of \cref{Eqr,Eqja1,Eqja2}, and considering $v_x=v_\perp{\rm sin}\theta$ and $v_y=v_\perp{\rm cos}\theta$, one gets:
\begin{align}
&\int_{-\infty}^tdt'e^{i(\vec{k}\cdot [\vec{r}'-\vec{r}] - \omega [t' - t])}
= \nonumber \\
&\sum_{\substack{m, m'\\n,'n}=-\infty}^\infty i^{(m'-m)}
\f{J_m\left(\f{-k_xv_\perp}{\Omega}\right) J_{m'}\left(\f{k_xv_\perp}{\Omega}\right)
J_n\left(\f{k_yv_\perp}{\Omega}\right) J_{n'}\left(\f{k_yv_\perp}{\Omega}\right)
}{i(k_zv_z + [n-m]\Omega - \omega)}e^{i(m'-m+n-n')\theta}.
\label{Eqexpint}
\end{align}

\item
\begin{align}
i\vec{k}\cdot\vec{v}'e^{i(\vec{k}\cdot\vec{r}'-\omega t')}
= \left(
i\omega + \f{d}{dt'} 
\right) e^{i(\vec{k}\cdot\vec{r}'-\omega t')}.
\label{Eqdt}
\end{align}

\end{enumerate}

Now, using \cref{Eqexpint,Eqdt}, the perturbed Fourier component of the distribution function in \cref{Eqf1} becomes:
\begin{align}
&\hat{\delta f}_{k_x} 
= -\frac{q}{\bar{T}} 
\Bigg\{ 
\hat{\delta \Phi}_{k_x} 
- \left( \sum_{k_x'} \hat{\omega}_{d, k_x - k_x'} \hat{\delta \Phi}_{k_x'} - \omega \hat{\delta \Phi}_{k_x} 
\right) \nonumber \\
&\sum_{\substack{m, m'\\n,'n} = -\infty}^{\infty}
i^{(m'-m)}
\f{ J_m\left(-\frac{k_x v_{\perp}}{\Omega}\right) J_{m'}\left(\frac{k_x v_{\perp}}{\Omega}\right) J_n\left(\frac{k_y v_{\perp}}{\Omega}\right) J_{n'}\left(\frac{k_y v_{\perp}}{\Omega}\right) 
} 
{k_z v_z + (n - m) \Omega - \omega}
e^{i(m' - m + n - n')\theta}
\Bigg\} f_0
\label{Eqdf}
\end{align}

Since we are only interested in low-frequency modes with $|\omega|\ll|\Omega|$, only the zeroth order cyclotron harmonic is considered in the rest of this derivation, implying $n=m$.\\

\subsubsection{Obtaining the dielectric function.}\label{SecDRDief}\hfill \vspace{1ex}

Taking the velocity integral of $\hat{\delta f}_{kx}$ in \cref{Eqdf}, one obtains the perturbed density $\hat{\delta N}_{kx} = \int d \vec{v}\ \hat{\delta f}_{kx}$:
\begin{align}
\delta \hat{N}_{k_x} = -\frac{N q}{T} 
\left\{ \hat{\delta \Phi}_{k_x} - \left( \sum_{k'_x} \hat{\omega}_{d,k_x - k'_x} \hat{\delta \Phi}_{k'_x} - \omega \hat{\delta \Phi}_{k_x} \right) 
\frac{1}{\omega} \left[ W\left( \frac{\omega}{|k_z| v_{th}} \right) - 1 \right]
Y(k_x\rho,k_y\rho),
\right\}
\label{Eqdn}
\end{align}
where, the resonant integral over $v_z$ can be expressed in terms of the dispersion function $W(z)=(1/\sqrt{2}\pi)\int_\Gamma dx\ x/(x-z)e^{-x^2/2}$ with the Landau integral path $\Gamma$ going from $-\infty$ to $\infty$ while avoiding the pole from below at $x=z$ to respect causality: 
\begin{align*}
\frac{1}{\omega} \left[ W\left( \frac{\omega}{|k_z| v_{th}} \right) - 1 \right]
= 
\f{1}{\sqrt{2\pi}}
\int \f{dv_z}{v_{th}}
\f{e^{-\f{1}{2}\f{v_z^2}{v_{th}^2}}}{k_zv_z - \omega},
\end{align*}
and the integral over the perpendicular velocity space, accounting for the finite Larmor radius effects is denoted by:
\begin{align*}
&Y(k_x\rho,k_y\rho)
= \\
&\int \f{v_\perp dv_\perp}{v_{th}^2} 
\sum_{n,n'=-\infty}^{\infty}
i^{n'-n}
J_{-n}\left(-\frac{k_x v_{\perp}}{\Omega}\right) J_{n'}\left(\frac{k_x v_{\perp}}{\Omega}\right) J_n\left(\frac{k_y V_{\perp}}{\Omega}\right) J_{n'}\left(\frac{k_y v_{\perp}}{\Omega}\right)
e^{-\f{1}{2}\f{v_\perp^2}{v_{th}^2}},
\end{align*}
where $kv_\perp/\Omega=k\rho\, v_\perp/v_{th}$

From the Poisson equation $-\nabla^2\delta \Phi=(1/\epsilon_0)\sum_{\rm species}q\delta N$, one finally obtains the dispersion relation in Fourier space:
\begin{align}
k^2\hat{\delta\Phi} 
- \f{1}{\epsilon_0}\sum_{\rm species}q\hat{\delta N}_{k_x}=0,
\label{Eqepsilon}
\end{align}
where $\hat{\delta N}_{k_x}$ has been defined in \cref{Eqdn}.

\subsubsection{Considering cosine corrugation to background pressure gradient.}\label{SecDRCos}\hfill \vspace{1ex}

A cosine variation to the background pressure gradient with a radial wavenumber $k_{x}^{\rm ext}$ is considered such that:
\begin{align}
\omega_d(x)
&=\hat{\omega}_{d,1}e^{-ik_{x}^{\rm ext}x} + \hat{\omega}_{d,0} + \hat{\omega}_{d,1}e^{ik_{x}^{\rm ext}x}
\nonumber \\
&=\hat{\omega}_{d,0} + \hat{\omega}_d^{\rm ext}~{\rm cos}(k_{x}^{\rm ext}x)
\label{Eqwdc}
\end{align}
where $\hat{\omega}_d^{\rm ext}=2\hat{\omega}_{d,1}$.\\
The electrostatic potential is also assumed to take the form:
\begin{align}
\delta \Phi
=\hat{\delta \Phi}_{-1}e^{-ik_{x}^{\rm ext}x} + \hat{\delta \Phi}_{0} + \hat{\delta \Phi}_{1}e^{ik_{x}^{\rm ext}x}.
\label{Eqphic}
\end{align}
Substituting \cref{Eqwdc,Eqphic} into \cref{Eqepsilon}, one obtains:
\begin{align}
\begin{bmatrix}
A_{-1,-1} & A_{-1,0} & 0\\
A_{0,-1} & A_{0,0} & A_{0,1}\\
0 & A_{1,0} & A_{1,1}
\end{bmatrix}
\begin{bmatrix}
\hat{\delta \Phi}_{-1}\\
\hat{\delta \Phi}_0\\
\hat{\delta \Phi}_1
\end{bmatrix}
=0,
\label{EqDRmat}
\end{align}
where the components of the dielectric coupling matrix $\bf A$ are:
\vspace{-2ex}
\begin{align*}
&A_{-1,-1}
=  1 + 
\sum_{s} \frac{1}{(k_1\lambda_D)^2}
\left\{1 + 
\frac{\omega - \hat{\omega}_{d,0}}{\omega}
\bigg[W
\bigg(\frac{\omega}{|k_z|v_{th}}
\bigg) - 1
\bigg]
Y(-k_x^{\rm ext}\rho,k_y\rho)
\right\},
\\
&A_{-1,0} 
=  
-\sum_{s} \frac{1}{(k_1\lambda_D)^2}
\left\{
\frac{\hat{\omega}_{d,1}}{\omega}
\bigg[W
\bigg(\frac{\omega}{|k_z|v_{th}}
\bigg) - 1
\bigg]
Y(-k_x^{\rm ext}\rho,k_y\rho)
\right\},
\\
&A_{0,-1} 
=  
-\sum_{s} \frac{1}{(k_0\lambda_D)^2}
\left\{
\frac{\hat{\omega}_{d,1}}{\omega}
\bigg[W
\bigg(\frac{\omega}{|k_z|v_{th}}
\bigg) - 1
\bigg]
Y(0,k_y\rho)
\right\},
\\
&A_{0,0} 
=  1 + 
\sum_{s} \frac{1}{(k_0\lambda_D)^2}
\left\{1 + 
\frac{\omega - \hat{\omega}_{d,0}}{\omega}
\bigg[W
\bigg(\frac{\omega}{|k_z|v_{th}}
\bigg) - 1
\bigg]
Y(0,k_y\rho)
\right\},
\\
&A_{0,1} 
=  
-\sum_{s} \frac{1}{(k_0\lambda_D)^2}
\left\{
\frac{\hat{\omega}_{d,1}}{\omega}
\bigg[W
\bigg(\frac{\omega}{|k_z|v_{th}}
\bigg) - 1
\bigg]
Y(0,k_y\rho)
\right\},
\\
&A_{1,0} 
=  
-\sum_{s} \frac{1}{(k_1\lambda_D)^2}
\left\{
\frac{\hat{\omega}_{d,1}}{\omega}
\bigg[W
\bigg(\frac{\omega}{|k_z|v_{th}}
\bigg) - 1
\bigg]
Y(k_x^{\rm ext}\rho,k_y\rho)
\right\},
\\
&A_{1,1} 
=  1 + 
\sum_{s} \frac{1}{(k_1\lambda_D)^2}
\left\{1 + 
\frac{\omega - \hat{\omega}_{d,0}}{\omega}
\bigg[W
\bigg(\frac{\omega}{|k_z|v_{th}}
\bigg) - 1
\bigg]
Y(k_x^{\rm ext}\rho,k_y\rho)
\right\},
\end{align*}
$(k_1)^2=(k_x^{\rm ext})^2+k_{y}^2+k_{z}^2$, $k_0^2=k_{y}^2+k_{z}^2$ and Debye length $\lambda_D=\epsilon_0 T/(Nq^2)$. The complex eigenfrequency $\omega$ can be obtained by solving ${\rm det}({\bf A})=0$.

In the limit of long wave-length such that $|k\lambda_D|\ll 1$, quasi-neutrality holds, and the first term, \emph{i.e.} `$1$', can be neglected in relation to the second term in $A_{-1,-1}, A_{0,0}$ and $A_{1,1}$ each. Furthermore, in case of adiabatic response for a species, \emph{i.e.} $\hat{\delta N}_{kx}=-(Nq/T)\hat{\delta \Phi}_{k_x}$ [see \cref{Eqdn}], only the term $1/(k\lambda_D)^2$ remains in each of the diagonal elements $A_{-1,-1}, A_{0,0}$ and $A_{1,1}$, and all the other elements, including the non-diagonal terms, can be set to zero.

\subsection{Dispersion relation results}\label{SecDRDerResult}\hfill

A solution to the dielectric function in \cref{EqDRmat} for an ETG case is presented in this subsection. The parameters considered are: $k_y\rho_e=0.1$, $k_x^{\rm ext}\rho_e=0.1$, $T_e=T_i$, $m_e/m_i=1/1836$, $1/L_{N}=d({\rm ln}N)/dx=10^{-5} \rho_e^{-1}$ and $1/L_{Te}=d({\rm ln}T_e)/dx=10^{-1}\rho_e^{-1}$. Quasi-neutrality, adiabatic ions and zero finite Larmor radius effects (\emph{i.e.} $Y=1$) are assumed. Only a corrugation in electron temperature is considered by setting the coefficient $\hat{\omega}_{Te}^{\rm ext}$ (in \cref{Eqwdc}); $\tilde{\omega}_d= \tilde{\omega}_N + \tilde{\omega_T}$, where, $\tilde{\omega}_N=(Tk_y/(qB)) d{\rm ln} N/dx$ and $\tilde{\omega}_T=(Tk_y/(qB)) d{\rm ln} T/dx$. All other corrugations are set to zero via $\hat{\omega}_{N}^{\rm ext}=0$ and $\hat{\omega}_{Ti}^{\rm ext}=0$. The resulting numerical solution is presented below.

Contour lines representing zeroes of the real (green lines) and imaginary (red lines) parts of the dielectric function $\epsilon(\omega)$ are plotted on the $\gamma-\omega$ complex plane in figure~\ref{FigDRCont}; $\gamma={\rm imag}(\omega)$ is the growth rate and $\omega_r={\rm real}(\omega)$ is the real frequency. The intersection of the green and red lines denotes the eigenfrequency of an ETG mode. The most unstable mode with the highest growth rate is shown. {\color{black}Note that negative real frequency corresponds to the electron diamagnetic drift direction.}

Figure~\ref{FigDRCont}(a) denotes the case with no corrugations, \emph{i.e.} $\hat{\omega}_{Te}^{\rm ext}=0$, and figure~\ref{FigDRCont}(b) denotes the case with a finite electron temperature corrugation with $\hat{\omega}_{Te}^{\rm ext}/\tilde{\omega}_{Te}=0.2$. These figures illustrate the mode splitting in the presence of finite cosine corrugation to the pressure gradient, whereby the original mode splits into three, with one being the original, one being more unstable and one being less unstable.

\begin{figure}[h] 
\centering
\includegraphics[scale=0.6]{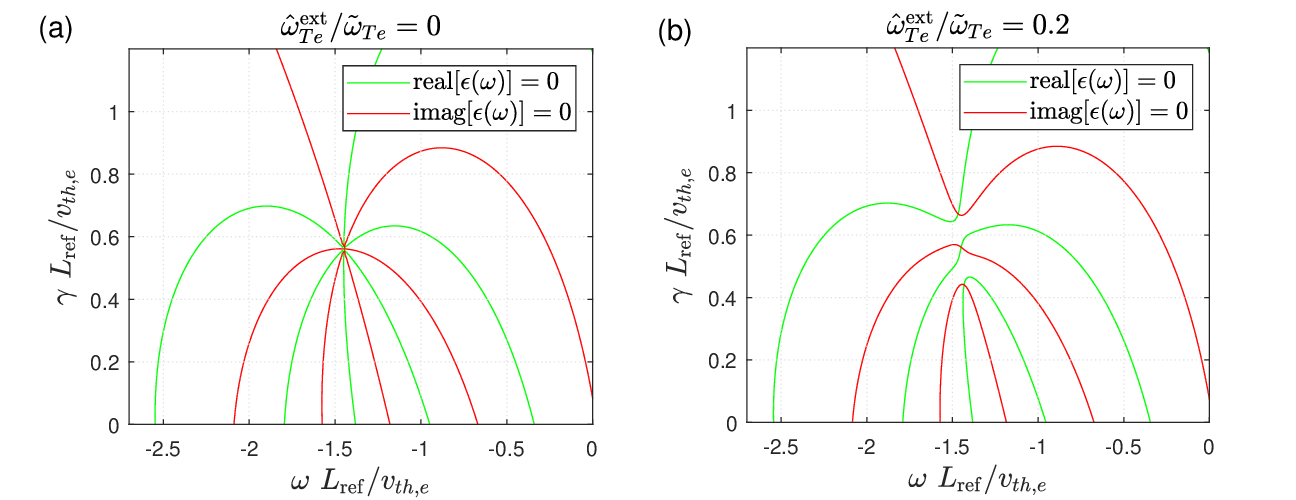}
\caption{Contour lines of zeroes of the real (green lines) and imaginary (red lines) parts of the dielectric function $\epsilon(\omega)$ plotted on the growth rate - real frequency plane. Intersection of green and red lines indicate a mode. (a) $\hat{\omega}_{Te}^{\rm ext}=0$, \emph{i.e.} no background $Te$ corrugation. (b) $\hat{\omega}_{Te}^{\rm ext}/\tilde{\omega}_{Te}=0.2$, \emph{i.e.} finite background $Te$ corrugation.}
\label{FigDRCont}
\end{figure}

In figure~\ref{FigDRGENEcomp}, the most unstable mode is plotted as a function of normalised background $T_e$ gradient $L_{\rm ref}/L_{Te}$ using dashed lines, assuming $L_{\rm ref}=1/k_z$, for three values of corrugation amplitudes $\hat{\omega}_{Te}^{\rm ext}/\tilde{\omega}_{Te}=0,\ 0.2$ and $0.5$. The growth rate increases and the real frequency decreases with $1/L_{Te}$ as expected for an ETG mode. Furthermore, the growth rate is found to increase with the corrugation amplitude. More analysis on this can be found on section~\ref{SecGENESlab} where a comparison of these results with the {\color{black}first-principle based gyrokinetic simulations} is made.

\begin{figure}[h] 
\centering
\includegraphics[scale=0.6]{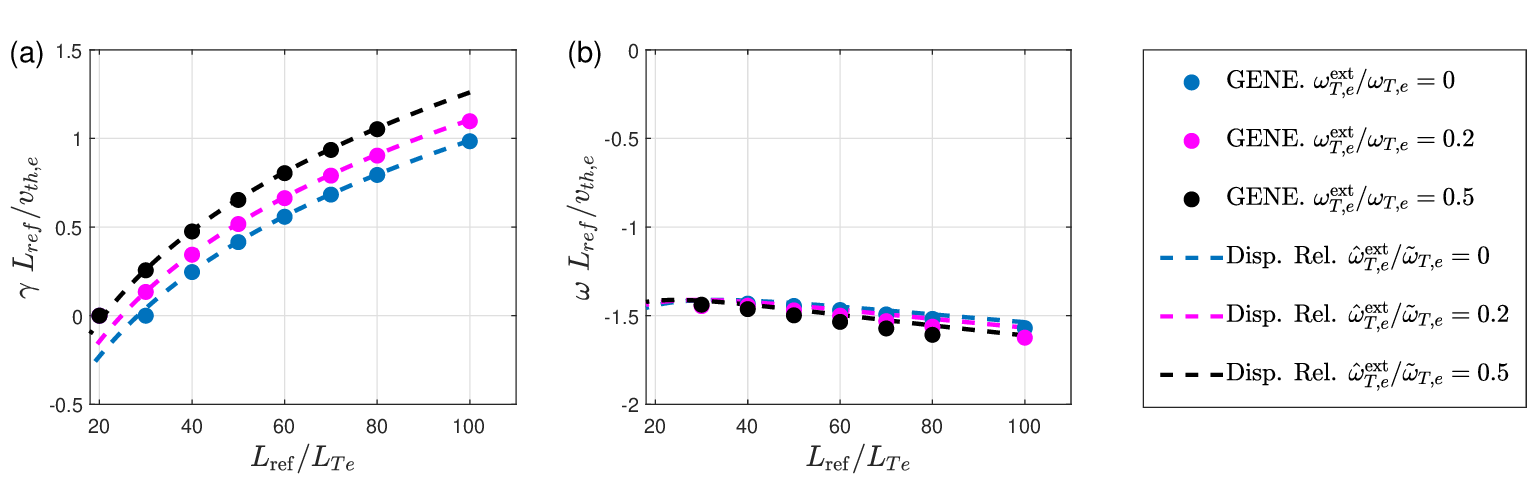}
\caption{(a) Growth rate $\gamma$ and (b) real frequency $\omega$ plotted vs normalised background $T_e$ gradient $L_{\rm ref}/L_{T_e}$. Dashed lines denote dispersion relation results and solid dots represent \textsc{GENE} slab results. Blue, magenta and black denote cases with $\hat{\omega}_{Te}^{\rm ext}/\tilde{\omega}_{Te}=\omega_{Te}^{\rm ext}/\omega_{Te}=0, 0.2$ and $0.5$ respectively.}
\label{FigDRGENEcomp}
\end{figure}

It is also possible to proceed analytically to solve the dispersion relation in \cref{EqDRmat}. In the limit of $k_x^{\rm ext}\ll k_y$, one can obtain the radial Fourier modes of the eigenmode structure as $\hat{\delta \Phi}_{-1}=\hat{\delta \Phi}_{1}=\pm \hat{\delta \Phi}_{0}/\sqrt{2}$, where $+$ and $-$ signs correspond to the more unstable and the less unstable modes respectively. The middle  branch has $\hat{\delta \Phi}_{0}=0$ and $\hat{\delta \Phi}_{-1}=\hat{\delta \Phi}_{1}$. The mode structure found by solving \cref{EqDRmat} numerically, as well as the gyrokinetic simulation in section~\ref{SecGENESlab}, agrees well with these results.



Given that ETG and ITG are homologous instabilities, one could obtain similar results presented in this section, including in figures~\ref{FigDRCont} and \ref{FigDRGENEcomp}, for the ITG modes as well.

\section{Gyrokinetic simulations with background $T_e$ corrugation}\label{SecGK}\hfill

In gyrokinetics, the fast gyromotion of charged particles around the background magnetic field lines is systematically eliminated, so that the 6 dimensional Vlasov-Maxwell system of equations becomes 5 dimensional, thereby saving computation time. Given that the microinstabilities and turbulence in fusion plasmas have much lower frequencies than the gyromotion frequency, the gyrokinetic formalism is justified.

The local version of the Eulerian gyrokinetic code \textsc{GENE}~\cite{GENE1} is used in this study. It uses a field aligned coordinate system~\cite{Beer1995}, where $x$ is the radial coordinate, $y$ the binormal coordinate and $z$ the parallel coordinate. Parallel velocity $v_\parallel$ and magnetic moment $\mu$ are the velocity coordinates.

A brief description of the implementation of background corrugations in \textsc{GENE} is given in this paragraph. The background density and temperature gradients responsible for the linear drive of instabilities manifest in the radial gradient of the background distribution function $f_0$. The latter appears as an advection ($v_{E,x}\nabla_xf_{0}$) with the $E\times B$ drift velocity ($\vec{v}_E=-\nabla\delta \Phi\times\vec{B_0}/B_0^2$) in the second term of the perturbed gyrokinetic equation as shown below for the electrostatic case:
\begin{align*}
\frac{\partial \delta f}{\partial t}
+ \vec{v}_E \cdot \left( \nabla f_{0} - \frac{\mu}{m v_{\|}} \nabla B_0 \frac{\partial f_{0}}{\partial v_{\|}} \right)
+ (\vec{v}_E + \vec{v}_{\nabla B} + \vec{v}_c) \cdot \vec{\Gamma}
+ v_{\|} \hat{b}_0 \cdot \vec{\Gamma}
- \frac{\mu}{m} \left( \hat{b}_0 + \frac{\vec{v}_c}{v_{\|} } \right) \cdot \nabla B_0 \frac{\partial \delta f}{\partial v_{\|}} = 0;
\end{align*}
$\vec{v}_{\nabla B}$ and $\vec{v}_c$ are the $\nabla B$ and curvature drift velocities respectively and $\Gamma=\nabla \delta f - (q/m v_\parallel)\nabla \delta \Phi \p f_0/\p v_\parallel$ (see equation~2.51 in \cite{LapillonnePhD} for more details).
The radial derivative of the background distribution function for the local Maxwellian considered in \textsc{GENE} is:
\begin{align}
\nabla_x f_0
=\left[
\frac{\omega_n}{L_{\rm ref}} + 
\left(
\f{mv_\parallel^2}{2T_0} + \f{\mu B_0}{T} - \f{3}{2}
\right) \frac{\omega_T}{L_{\rm ref}}
-\f{\mu}{T_0}\f{\p B_0}{\p x}
\right]f_0,
\end{align}
where $\omega_n=L_{\rm ref}\,d{\rm ln} N/dx$ and $\omega_T=L_{\rm ref}\,d{\rm ln} T/dx$ are the usual (radially constant) drive gradients of the background density and temperature respectively. A sinusoidal radial corrugation to the background temperature gradient is added by modifying the term $\omega_T$ appearing in $v_{E,x}\nabla_x f_0$, \emph{i.e.}:
\begin{align}
\omega_T \longrightarrow \underbrace{\omega_T + \omega_T^{\rm ext}{\rm sin}[(2\pi/L_x)k_{x,{\rm ind}}x + \varphi^{\rm ext}]}_{\omega_T^{\rm tot}},
\label{Eqwtot}
\end{align}
where $\omega_T^{\rm ext}$ is the amplitude of the external corrugation, $k_{x,{\rm ind}}$ is the wave-number and $\varphi^{\rm ext}$ is the phase. In all simulations considered in this work, scans in $\varphi^{\rm ext}$ have revealed that it has little effect on the results and is set to $\pi$.

\subsection{Linear slab-ETG with zero shear}\label{SecGENESlab}\hfill

Linear initial-value simulation results for slab-ETG is given in this subsection to verify the dispersion relation results in section~\ref{SecDRDerResult}. 

Slab geometry is considered with zero magnetic shear. The grid resolutions used are $N_x\times N_z\times N_{v_\parallel}\times N_\mu=4\times 32\times 32\times 8$. Simulation box-widths are $L_x=62.8\rho_e$,  $L_{v_\parallel}=3\sqrt{2}~v_{th,e}$ and $L_{\mu}=9~T_e/B_{0,\rm axis}$. Both numerical resolution and box-widths have been chosen after convergence tests. The $y-$wavenumber chosen is $k_y\rho_e=0.1$ and the background density gradient is $\omega_N=0$. For the default case, the background electron temperature gradient is $\omega_{Te}=60$. Fully electrostatic limit is taken by setting $\beta=0$, and ions are assumed to respond adiabatically. {\color{black} Unlike in the toroidal geometry, the parallel direction is periodic in the slab case and the radial Fourier wavenumbers $k_x$s are linearly independent. The parallel box-length is $L_z=2\pi L_{\rm ref}$, where $L_{\rm ref}$ is the reference length.} The radial wavenumber of the background corrugation considered is $k_{x,\rm ind}=1$, \emph{i.e.}, one sinusoidal variation in background electron temperature is considered in the radial domain, with a maximum to the left of $x=0$ and a minimum to the right of $x=0$ as described in equation~(\ref{Eqwtot}). In all these simulations, the modes are found to have one sinusoidal variation in $L_z$, meaning $k_z=1/L_{\rm ref}$, and therefore the results can be compared with the dispersion relation results in figure~\ref{FigDRGENEcomp}, for which $k_z=1/L_{\rm ref}$ assumption has been made.

The growth rate $\gamma$ and real frequency $\omega_r$ of the most unstable modes are plotted as a function of the normalised radially constant background $T_e$ gradient $L_{\rm ref}/L_{Te}$ in figure~\ref{FigDRGENEcomp} using dots. Three different values of background $T_e$ corrugation amplitudes $\omega^{\rm ext}_{Te}$ are considered: $\omega^{\rm ext}_{Te}/\omega_{Te}=0, 0.2$ and $0.5$. Indeed, as expected for an ETG mode, the growth rate increases and the real frequency decreases with $L_{\rm ref}/L_{Te}$. Furthermore, as predicted by the dispersion relation analysis, the modes become more unstable with increasing background $T_e$ corrugation amplitude. The mode structure also matches well with dispersion relation result. The dispersion relation results, which are based on a number of simplifications and assumptions, show a fairly good match with the more comprehensive first-principle based \textsc{GENE} simulations.

{\color{black} The mode structure of the electrostatic potential in the $(x,z)$ plane  is shown in figure~\ref{Figsetgmodestruct}. A higher radial resolution of $N_x=32$ is considered for these simulations. The mode is radially uniform for the case without background corrugations, whereas for the case with finite background corrugations, the mode can be found to localize near the radial position of maximum electron temperature gradient.}

\begin{figure}[h!] 
\centering
\includegraphics[scale=0.6]{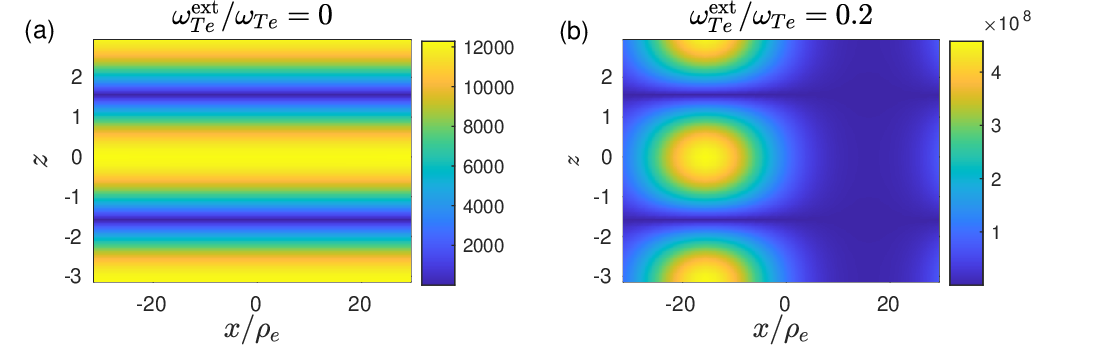}
\caption{Electrostatic potential $\delta\Phi$ plotted on the $x-y$ plane for the most unstable ETG modes in linear \textsc{GENE} slab simulations for the case with (a) no background temperature corrugation, \emph{i.e.} $\omega^{\rm ext}_{Te}/\omega_{Te}=0$, and (b) a finite background temperature corrugation of $\omega^{\rm ext}_{Te}/\omega_{Te}=0.2$.}
\label{Figsetgmodestruct}
\end{figure}

\subsection{Linear toroidal-ETG}\label{SecLin}\hfill

The toroidal-ETG case is inspired from the ETG benchmark paper~\cite{Nevins2006}. `s-alpha' geometry is considered with a safety factor of $q_0=1.4$, {\color{black} magnetic shear of $\hat{s}=(r/q_0)dq/dr=1$}, and $r/R=0.18$, where $r$ is the radial position of the fluxtube and $R$ is the major radius of the tokamak. A binormal wavenumber of $k_y\rho_e=0.4$ is taken as the default case. The numerical parameters used are $N_x\times N_z\times N_{v_\parallel}\times N_\mu=32\times 16\times 32\times 8$. In addition to $L_x=1/(\hat{s}k_y)=2.5\rho_e$ (where $1/(\hat{s}k_y)$ is the distance between mode rational surfaces), a larger box-size of $L_x=20\rho_e$ is also chosen so that it is more comparable to that in the nonlinear simulations in section~\ref{SecNonlin}. The results remain converged for larger box-sizes. {\color{black}$L_z=2\pi$ and $L_{\rm ref}=R$.} A background temperature gradient of $\omega_{Te}=20$ and density gradient of $\omega_{Te}=2.2$ are chosen.

Growth rate as a function of the background gradient $\omega_{Te}$ is plotted in figure~\ref{Figgwvsomte}, showing the critical gradient at $\omega_{Te}=5.5$. The $k_y$ spectra are plotted in figure~\ref{Figgwvsky}. The modes are linearly unstable for $k_y\rho_e>2$. However in nonlinear simulations, only the minimum $k_y$ mode considered ($k_y\rho_e=0.4$ for the default case) remains dominant, as will be discussed in more detail in the next subsection.

\begin{figure}[h!] 
\centering
\includegraphics[scale=0.6]{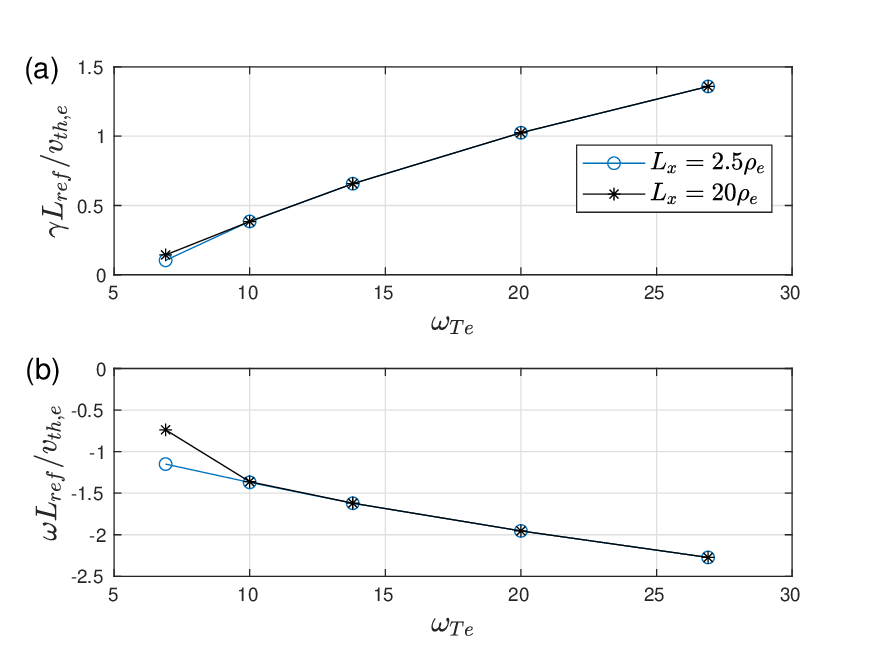}
\caption{(a) Growth rate $\gamma$ and (b) real frequency $\omega$ for linear toroidal ETG \textsc{GENE} runs plotted as a function of the normalised background $T_e$ gradient $\omega_{Te}$. Blue and black denote radial box-widths of $L_x=2.5\rho_e$ and $L_x=20\rho_e$ respectively.}
\label{Figgwvsomte}
\end{figure}

\begin{figure}[h!] 
\centering
\includegraphics[scale=0.6]{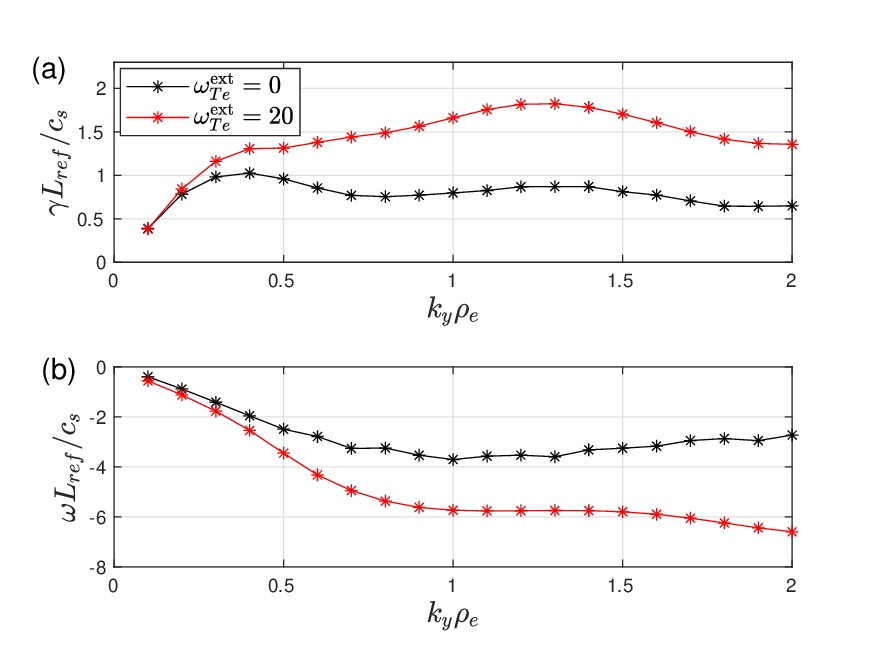}
\caption{(a) Growth rate $\gamma$ and (b) real frequency $\omega$ for linear toroidal ETG \textsc{GENE} runs plotted as a function of the binormal wavenumber $k_y$. Black and red denote cases with no corrugation and a corrugation amplitude of $\omega_{Te}^{\rm ext}=20$.}
\label{Figgwvsky}
\end{figure}

The growth rate increases with the corrugation amplitude $\omega_{Te}^{\rm ext}$ as shown in figure~\ref{Figgwvsox}, similar to the slab-ETG result in section~\ref{SecGENESlab}. 

\begin{figure}[h!] 
\centering
\includegraphics[scale=0.6]{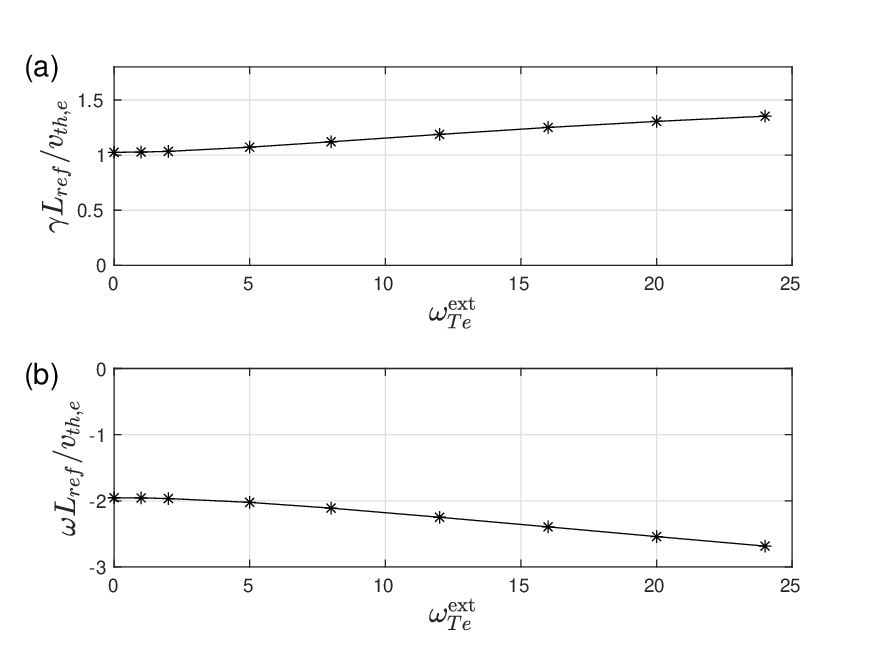}
\caption{(a) Growth rate $\gamma$ and (b) real frequency $\omega$ for linear toroidal ETG \textsc{GENE} runs plotted as a function of the corrugation amplitude $\omega_{Te}^{\rm ext}$.}
\label{Figgwvsox}
\end{figure}

\subsection{Nonlinear toroidal-ETG}\label{SecNonlin}\hfill

While linear simulations can predict the growth-rate of the modes, it does not give a complete picture of the turbulent transport they produce. Various nonlinear saturation mechanisms play an important role in predicting the particle and heat flux in experiments, and therefore nonlinear simulations become relevant.

\subsubsection{Simulation set-up}\hfill \vspace{1ex}

{\color{black} Unless specified otherwise, the same parameters as in the linear simulations in section~\ref{SecLin} are considered for the nonlinear simulations. A background electron temperature gradient of $\omega_{Te}=20$ and a magnetic shear of $\hat{s}=1$ considered for the linear simulations are more typical of the pedestal, and are labeled \emph{case 1} for the nonlinear simulations. For reference, an additional set of parameters, labeled \emph{case 2}, is also considered with lower values of $\omega_{Te}=6.9$ and $\hat{s}=0.1$ that are more typical of the core. See table~\ref{TablePar} for the corresponding numerical parameters for the two cases. No collisions are considered unless specified otherwise. A fourth-order numerical hyper-diffusion term (-$\nu_z\partial^4/\partial z^4$) is added for the parallel coordinate on the right hand side of the gyrokinetic equation, and the hyper-diffusion coefficient $h_z$ is defined as $h_z=\nu_z/\Delta z^4$ ~\cite{Pueschel2010_2}. The simulations are converged in all the chosen numerical parameters except for the two instances addressed in the following two paragraphs; however, they do not affect the physics results and conclusions.}

\begin{table}[h!]
\centering
\begin{tabular}{ccc}
\textbf{Parameter} & \textbf{Case 1} & \textbf{Case 2} \\ \hline
$\omega_{Te}$      & 20      & 6.9      \\ 
$\hat{s}$      & 1      & 0.1      \\ 
$k_{y,\min}$      & 0.4      & 0.1      \\
$L_x \times L_y$      & $400\rho_e \times 15.7 \rho_e$      & $400\rho_e \times 62.8 \rho_e$      \\
$N_x\times N_{y}$      & $160\times 12$      & $200\times 16$      \\ 
$h_z$      & 2      & 0.5      \\ 
\end{tabular}
\caption{The two parameter sets considered for the nonlinear simulations.}
\label{TablePar}
\end{table}

{\color{black} In these simulations, most transport is carried by $k_y=k_{y,\min}$ and the fluxes are not yet converged in $k_{y,\min}$ even for an order of magnitude lower value. The same issue has been reported in the ETG benchmark reference~\cite{Nevins2006}. One could assume that a flux-tube simulation with a chosen $k_{y,\min}$ amounts to modeling the full flux-surface  (corresponding to a minimum toroidal mode number of $n_{\min}=1$) of a tokamak with $\rho^\star=k_{y,\min}\rho_i(r_0/a)(1/q_0)$, where $\rho^\star=\rho_i/a$ measures the system size; see references~\cite{Ball2019} and \cite{AjayCJ2020} for more details.}

{\color{black} For the case 1 set of parameters, when finite background temperature corrugations are considered, the fluxes are found to be sensitive to the radial resolution. This may be a result of the lower fluxes in presence of finite background corrugations (as will be discussed in detail in the following subsection) and the simulation being closer to marginality, where such a sensitivity to numerical resolutions can be expected. Nevertheless, compared to the case with zero background corrugations, a significant reduction in flux is seen when finite background corrugations are considered, irrespective of the radial resolution.}


\subsubsection{Reduced transport and profile shearing}\label{SebsebNLResult}\hfill \vspace{1ex}

In presence of finite corrugations on the background $T_e$, a decrease in heat flux is observed. This is shown in figure~\ref{FigQvstoxcomp} for the two cases considered. This is particularly interesting given that the maximum linear growth rate increases with finite corrugations as discussed in sections~\ref{SecDRDerResult}, \ref{SecGENESlab} and \ref{SecLin} and one would have expected an increase in flux instead. The reduction in fluxes in presence of background pressure corrugation is in line with the findings of references~\cite{Li2024,Candy2025}, where the stabilising effect of profile curvature has been reported. In this subsection, the saturation mechanism, \emph{i.e.} profile shearing, that may be responsible for the lower fluxes in cases with finite background corrugations is discussed. 

\begin{figure}[h!] 
\centering
\includegraphics[scale=0.6]{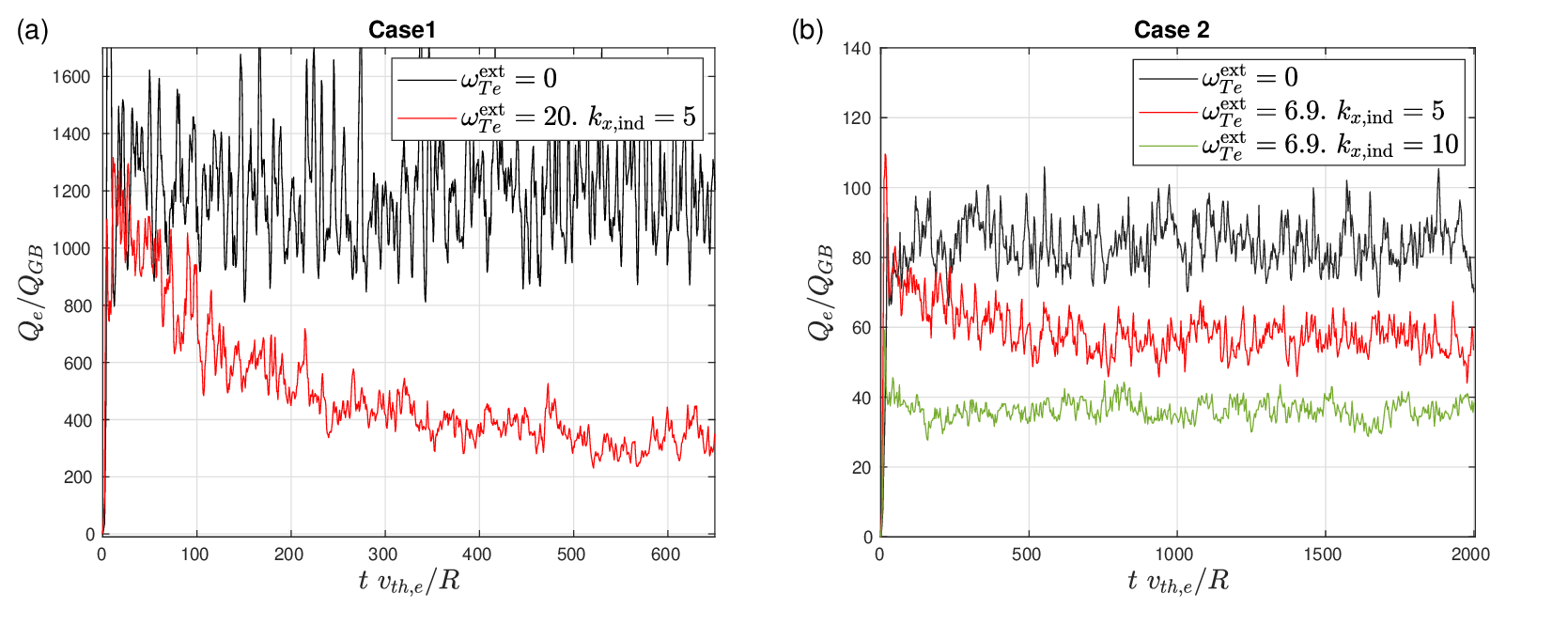}
\caption{Gyro-Bohm normalised electron heat flux plotted as a function of normalised time $t\ v_{th,e}/R$ for (a) case 1 ($\omega_{Te}=20$ and $\hat{s}=1$) and (b) case 2 ($\omega_{Te}=6.9$ and $\hat{s}=0.1$). Black lines denote simulations with no background corrugation. Red and green denote simulations with a background corrugation amplitude of $\omega_{Te}^{\rm ext}=\omega_{Te}$ and a corrugation wavenumber of $k_{x,\rm ind}= 5$ and $10$ respectively.}
\label{FigQvstoxcomp}
\end{figure}

Many of the microinstabilities, including ETG and ITG, have real frequencies $\omega_r$ that are proportional to the diamagnetic frequency $\omega_{\rm dia}=-\nabla P\times\vec{B_0}/(qB_0^2)$. When the background pressure gradient $\nabla P$ varies with the radius, so does $\omega_{\rm dia}$. The real frequency $\omega_r$ may also therefore vary with the radius. This is verified in figure~\ref{Figwvsx}, where the real frequencies measured from electrostatic potential fluctuations $\delta \Phi$ in nonlinear simulations are plotted as a function of the radial coordinate. {\color{black}Assuming that a single dominant mode having a time dependence of $e^{i\omega t}$ governs the dynamics of fluctuating quantities, say of $\delta \Phi$, then the dominant real frequency is found such that $\omega_r={\rm Imag}({\rm ln}[\delta \Phi(t+\Delta t)/\delta \Phi(t)])/\Delta t$. The simulation time-window considered for obtaining the frequencies is $tv_{th,e}/R=20-100$. The diamagnetic frequency $\omega_{\rm dia}R/v_{th,e}=k_y\rho_e\omega_T^{\rm tot}$ is divided by a factor of $12.8$ to match scales with the real frequency plots; $\omega_T^{\rm tot}$ is the total background gradient including the corrugation [see \cref{Eqwtot}].}  For the simulation with finite corrugation, the real frequency is indeed closely proportional to the variation of the diamagnetic frequency. 

\begin{figure}[h!] 
\centering
\includegraphics[scale=0.6]{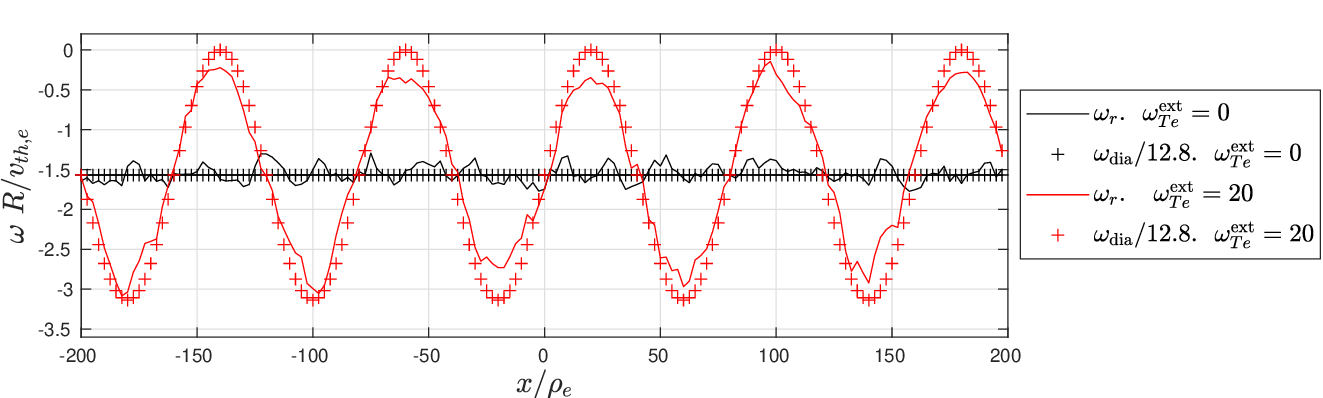}
\caption{Solid lines denote the real frequency $\omega_r$ measured from electrostatic potential at each radial position $x$ in nonlinear \textsc{GENE} simulations. Black denotes the case with no corrugation. Red denotes the case with a corrugation amplitude of $\omega_{Te}^{\rm ext}=20$ and a corrugation wavenumber of $k_{x,\rm ind}=5$. The `plus' markers denote the diamagnetic frequency $\omega_{\rm dia}R/v_{th,e}=k_y\rho_e\omega_T^{\rm tot}$ divided by a scaling factor.}
\label{Figwvsx}
\end{figure}

The phase velocity ($\omega_r/k_y$) of the waves associated with the modes also therefore varies with the radius. Such a radial variation in the phase velocity can shear and break the turbulent eddies, in a process that is homologous with $E\times B$ zonal flow shearing~\cite{Biglari1990,Rosenbluth1998}, therefore lowering turbulent fluxes.

The profile shearing of turbulent eddies can be seen in figure~\ref{Figphionxandz}, where the snapshots of time evolution of the electrostatic potential on the $x-y$ plane at the outboard mid-plane ($z=0$) is shown for three different times. Note that, a zoom of the radial domain at $x=0$ corresponding to one wavelength ($=400/k_{x,\rm ind}$, where $k_{x,\rm ind}=5$) of the sinusoidal background corrugations is shown; for the chosen values of $\omega_{Te}=20$, and $\omega_{Te}^{\rm ext}=20$, the total background gradient $\omega_{Te}^{\rm tot}$ reaches $40$ on the left of $x=0$ and $0$ on the right of $x=0$. In the case without corrugations, the turbulent eddies move up along $y$, in the electron diamagnetic direction, all through the radial domain with similar velocity. Whereas in the case with corrugation, higher velocities are seen on at the radial positions corresponding to higher total background gradient $\omega_{Te}^{\rm tot}$, \emph{i.e.} to the left of $x=0$, and the eddies can be seen to be near stationary in the radial locations corresponding to lower $\omega_{Te}^{\rm tot}$ ($\simeq 0$), \emph{i.e.} to the right of $x=0$. The radial eddy widths are also lower in the case with finite corrugation, consistent with the lower flux value.

\begin{figure}[h!] 
\centering
\includegraphics[scale=0.6]{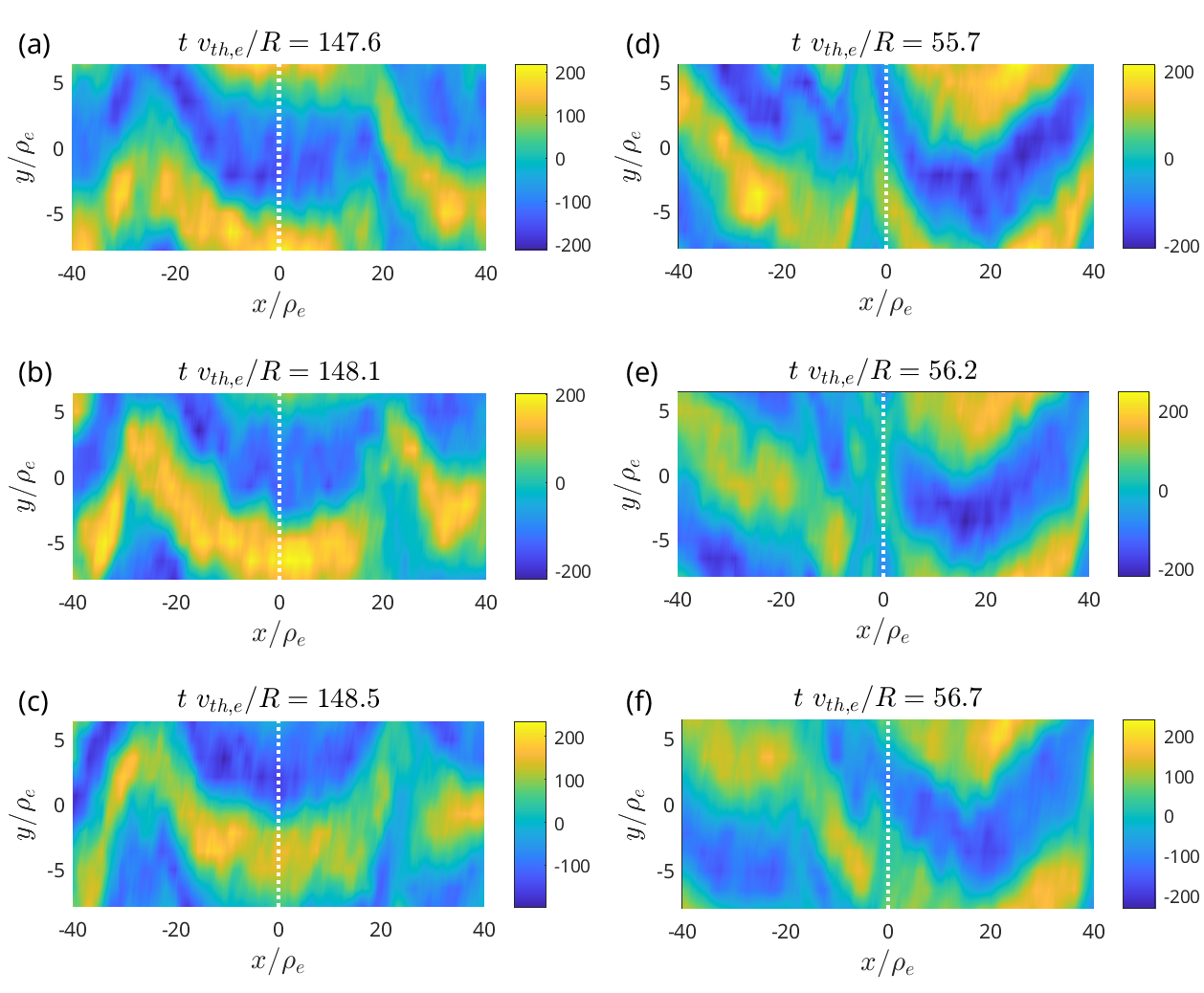}
\caption{Electrostatic potential plotted on the $y-x$ plane in  nonlinear \textsc{GENE} simulations with (left, a-c) no corrugation and (right, d-f) with a corrugation amplitude of $\omega_{Te}^{\rm ext}=20$ and corrugation wavenumber of $k_{x,\rm ind}=5$, each at three different snapshots in time.}
\label{Figphionxandz}
\end{figure}

{\color{black}To further examine the effect of profile shearing, the results of a scan in the corrugation wavenumber is also included in figure~\ref{FigQvstoxcomp}(b). With higher corrugation wavenumber, \emph{i.e.} shorter radial shearing length, the turbulent eddy width decreases and lower fluxes are obtained. }

{\color{black} In the ETG simulations having corrugated background electron temperature gradient, the instability drive also varies with radius, and one could wonder if it plays a role in explaining the reduced fluxes vis-\`a-vis the profile shearing mechanism. Given that the diamagnetic drift frequency $\omega_{\rm dia}$ is also a function of the density gradient, one can isolate the effect of profile shearing by considering only finite background density gradient corrugation and no background temperature gradient corrugation. The result of such a scan in only the background density corrugation amplitude $\omega^{\rm ext}_N$ is shown in figure~\ref{FigQvstoxncomp}. With increasing amplitude of density corrugations, effect of profile shearing intensifies and flux decreases, confirming that profile shearing is indeed the dominant saturation mechanism.}

\begin{figure}[h!] 
\centering
\includegraphics[scale=0.6]{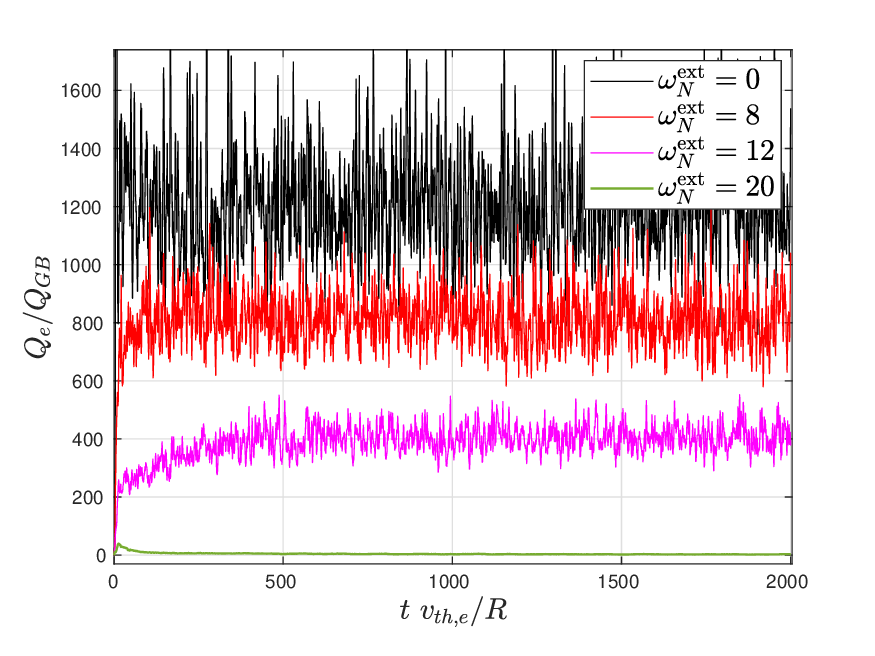}
\caption{Gyro-Bohm normalised electron heat flux plotted as functions of normalised time $t\ v_{th,e}/R$ for case 1 set of parameters. Black line denotes simulation with no background corrugation. Red, magenta and green denote simulations with a background density corrugation amplitude of $\omega_{N}^{\rm ext}=8, 12$ and $20$ respectively, and a corrugation wavenumber of $k_{x,\rm ind}=5$.}
\label{FigQvstoxncomp}
\end{figure}


Collisions are found to have little effect on flux levels in simulations without background corrugations. A scan on electron collisionality $\nu_e^*$ (=collision freq./bounce freq.) in case 1 simulations with finite background corrugations shows that the gyro-Bohm normalised fluxes $Q_e/Q_{GB}$ increases gradually with increasing collisionality, reaching a value of $Q_e/Q_{GB}\simeq 820$ at $\nu_e^*=27.6$, which is a typical value of collisionality at the pedestal bottom; $Q_e/Q_{GB}\simeq 590, 680$ and $820$ for $\nu_e^*=0.3, 2.8$ and $27.6$ respectively [see figure~\ref{FigQvstoxcomp}(a) for reference]. $Q_e/Q_{GB}\simeq 820$ is still 32\% less than $Q_e/Q_{GB}\simeq 1200$ in simulations without background corrugations, \emph{i.e.} the effect of profile shearing is significant even in presence of finite collisionality.

\subsubsection{{\color{black} Effect of zonal structures}}\label{SebsebNLResult} \hfill \vspace{1ex}

{\color{black} Stationary zonal structures play an important role in the dynamics and saturation of ion-scale turbulence and often tend to flatten or steepen the background gradients~\cite{Waltz2006,Dominski2015,Justin2020,Hatch2021,AjayCJ2023,Volcokas2025}. In the absence of any background corrugations, ETG turbulence (with adiabatic ion response) does not exhibit any such stationary zonal structures. However, in presence of background corrugations, the perturbed zonal quantities of temperature, density, electrostatic potential etc. respond to the corrugation. In figure~\ref{Figstaircase}, their time-averaged profiles are plotted as a function of the radial coordinate. The effective electron temperature gradient $\omega_{Te}^{\rm eff}$ is defined as $\omega_{Te}^{\rm eff}=\omega_{Te}^{\rm tot} + (R/T_{0,e})\,d\langle \delta T_e\rangle/dx$, where $\langle\cdot\rangle$ is the flux-surface and time average, such that it denotes the net effective temperature gradient seen by the turbulence including the background corrugation [see equation~(\ref{Eqwtot}) for definition of $\omega_{Te}^{\rm tot}$] and the zonal perturbed electron temperature contribution. Note that the zonal response negates the background corrugation but not fully. The effective density gradient $\omega_N^{\rm eff}$, defined as $\omega_{Te}^{\rm eff}=\omega_{Te}^{\rm tot} + (R/N_0)\,d\langle \delta N\rangle/dx$, and zonal flow shearing rate $\omega_{E\times B}$ defined as $\omega_{E\times B}=(1/B_0)d^2\langle\delta \phi\rangle/dx^2$ also exhibit significant corrugations. Artificially switching off zonal flows in these simulations is found to reduce the fluxes, which indicates that they do not play an active role in saturation~\cite{Lin1998}.}

\begin{figure}[h!] 
\centering
\includegraphics[scale=0.6]{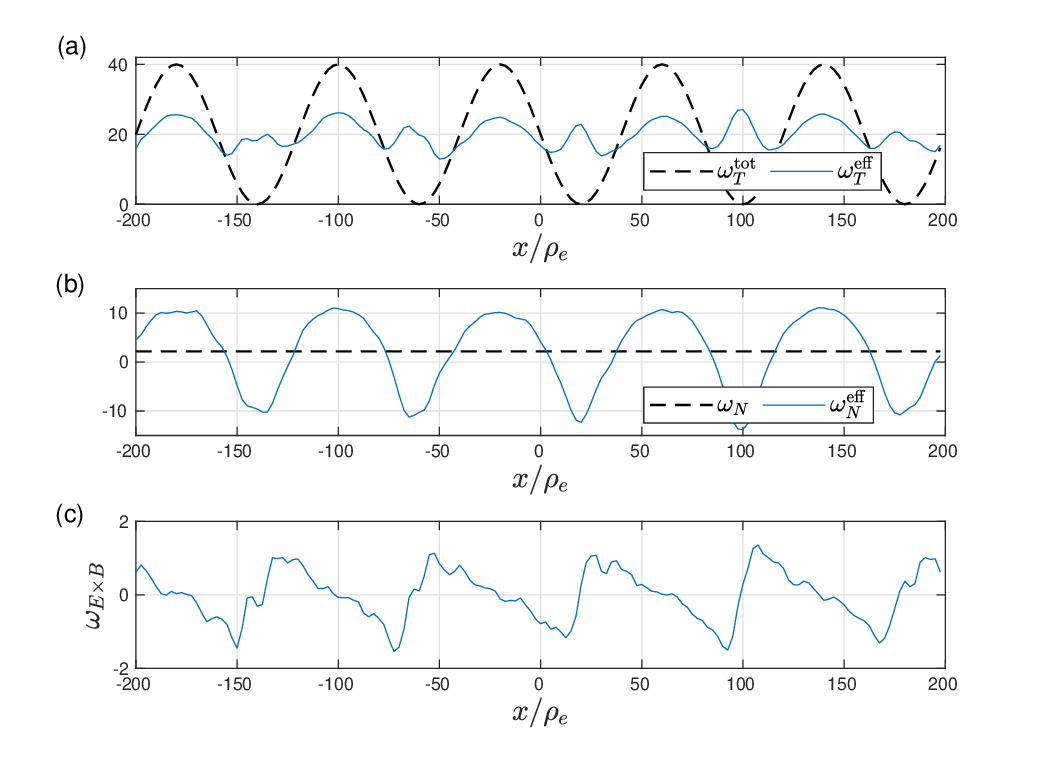}
\caption{Radial profile of zonal structures in case 1 simulation with a background temperature corrugation amplitude $\omega_{Te}^{ext}=20$ and a corrugation wavenumber $k_{x,\rm ind}=5$. (a) Black dotted line denotes the total background temperature gradient and blue line denotes the effective temperature gradient $\omega_{Te}^{\rm eff}$ that includes the effect of time averaged zonal temperature perturbation. (b) Black line denotes the background density gradient $\omega_N$ and blue line denotes the effective density gradient $\omega_{Te}^{\rm eff}$. (c) Zonal flow shearing rate $\omega_{E\times B}$.}
\label{Figstaircase}
\end{figure}

{\color{black} The case 1 simulations with background electron temperature corrugations, when extended for longer times [beyond $t v_{th,e}/R=700$, in figure~\ref{FigQvstoxcomp}(a)], exhibit development of prominent stationary zonal structures with very narrow radial widths of the order of few electron Larmor radii, and an associated drop in fluxes. However, in a tokamak, such super fine structures are unphysical because of effects such as collisions. In presence of finite collisionality, these super fine scale structures and the associated sudden drop in fluxes are indeed found to be absent.} 

\vfill

\section{Conclusions}\label{SecConc}\hfill

The dependence of ETG modes on background pressure corrugations was discussed in this paper. Such background corrugations could result from microinstabilities~\cite{Waltz2006,Dominski2015,AjayCJ2020,Hatch2021,AjayCJ2023}, such as for instance microtearing modes~\cite{Hatch2021,AjayCJ2023} in the pedestal that can almost completely flatten $T_e$ over radial widths of the order of a few ion Larmor radii.

A simple sinusoidal corrugation on the background $T_e$ gradient was considered to analyse its consequence on ETG modes. It was shown, both with the help of a slab dispersion relation analysis and linear gyrokinetic simulations, that the linear growth rate of an ETG mode increases with increasing amplitude of background $T_e$ corrugation. However in nonlinear simulations, the profile shearing~\cite{Waltz2002} resulting from the background corrugations was found to break and radially decorrelate the turbulent eddies, reducing the fluxes to a roughly a third. This was validated by plotting the real frequency ($\propto$ phase velocity) and showing that it indeed follows the radial variation of the diamagnetic frequency which results from the background pressure corrugations.

{\color{black}Microtearing and ETG are relevant microinstabilities in the pedestal. Therefore, a suppression of ETG transport resulting from ion Larmor radius scale pressure corrugations caused by microtearing modes is important for pedestal formation and L-H mode transitions.} However, turbulence suppression via profile-shearing is in principle a universal phenomenon valid all throughout the tokamak and for most microinstabilities at varying degrees. This is corroborated by the (increase in linear growth rate and) decrease in fluxes observed in preliminary ITG simulations with a corrugated background pressure.

Experimentally measuring the fine-scale corrugations remains a challenge. Furthermore, usually the experimental data is smoothed and fitted before it is used as an input for both both gyrokinetic and in reduced turbulence models such as quasi-linear models in integrated modeling simulations, which could lead to an over-prediction of transport. Properly including the effect of this profile shearing is therefore necessary for realistic transport predictions.

{\color{black}In addition to profile shearing, another mechanism that may be relevant and could be investigated in the future is the possibility of improved saturation efficiency due to coupling of the unstable mode with subdominant or stable modes~\cite{Hatch2011_2,MakwanaPhDthesis,Terry2014}. Given the mode splitting in presence of background corrugation, the mode density increases, thereby increasing the probability for a higher triplet correlation time that facilitates energy transfer to damped modes.}

\section*{Acknowledgments}
This work has been carried out within the framework of the EUROfusion Consortium, funded by the European Union via the Euratom Research and Training Programme (Grant Agreement No 101052200 — EUROfusion). Views and opinions expressed are however those of the author(s) only and do not necessarily reflect those of the European Union or the European Commission. Neither the European Union nor the European Commission can be held responsible for them.

\bibliography{ETGCorrugations} 
\bibliographystyle{unsrt}

\end{document}